\documentclass[12pt]{article}
\usepackage{amssymb,epsf}
\def\lsim{\mathrel{\rlap {\raise.5ex\hbox{$ < $}}
{\lower.5ex\hbox{$\sim$}}}}
\def\gsim{\mathrel{\rlap {\raise.5ex\hbox{$ > $}}
{\lower.5ex\hbox{$\sim$}}}} 
\def\sqr#1#2{{\vcenter{\vbox{\hrule height.#2pt

        \hbox{\vrule width.#2pt height#1pt \kern#1pt

           \vrule width.#2pt}

        \hrule height.#2pt}}}}

\def\lsim{{\displaystyle
{{\raise-8pt\hbox{$ <$}}
\atop{\raise5pt\hbox{$\sim$}}}}}
\def\gsim{{\displaystyle
{{\raise-8pt\hbox{$ >$}}
\atop{\raise5pt\hbox{$\sim$}}}}}
\def\slsim{{\displaystyle
{{\raise-8pt\hbox{$\scriptstyle <$}}
\atop{\raise5pt\hbox{$\scriptstyle \sim$}}}}}
\def\sgsim{{\displaystyle
{{\raise-8pt\hbox{$\scriptstyle  >$}}
\atop{\raise5pt\hbox{$\scriptstyle \sim$}}}}}
\newskip\humongous \humongous=0pt plus 1000pt minus 1000pt

\newcommand{\sumpf}[0]{\sum_{(H^{\rm f},G^{\rm f})}\! \! \! \!
{\raise
4pt
\hbox{$'$}}\,}

\newcommand{\sump}[0]{\sum_{(H,G)}\! \! {\raise 4pt \hbox{$'$}}\,}

\def\bs{\begin{subequations}}
\def\es{\end{subequations}}
\catcode`\@=11
\newcount\hour
\newcount\minute
\newtoks\amorpm
\hour=\time\divide\hour by60
\minute=\time{\multiply\hour by60 \global\advance\minute by-\hour}
\edef\standardtime{{\ifnum\hour<12 \global\amorpm={am}%
        \else\global\amorpm={pm}\advance\hour by-12 \fi
        \ifnum\hour=0 \hour=12 \fi
        \number\hour:\ifnum\minute<10 0\fi\number\minute\the\amorpm}}
\edef\militarytime{\number\hour:\ifnum\minute<10 0\fi\number\minute}
\def\draftlabel#1{{\@bsphack\if@filesw {\let\thepage\relax
   \xdef\@gtempa{\write\@auxout{\string
      \newlabel{#1}{{\@currentlabel}{\thepage}}}}}\@gtempa
   \if@nobreak \ifvmode\nobreak\fi\fi\fi\@esphack}
        \gdef\@eqnlabel{#1}}
\def\@eqnlabel{}
\def\@vacuum{}
\def\draftmarginnote#1{\marginpar{\raggedright\scriptsize\tt#1}}
\def\draft{\oddsidemargin -.2truein
        \def\@oddfoot{\sl preliminary draft \hfil
        \rm\thepage\hfil\sl\today\quad\militarytime}
        \let\@evenfoot\@oddfoot \overfullrule 3pt
        \let\label=\draftlabel
        \let\marginnote=\draftmarginnote
   \def\@eqnnum{(\theequation)\rlap{\kern\marginparsep\tt\@eqnlabel}%
\global\let\@eqnlabel\@vacuum}  }
\def\subequations{\refstepcounter{equation}%
  \edef\@savedequation{\the\c@equation}%
  \@stequation=\expandafter{\theequation}
  \edef\@savedtheequation{\the\@stequation}
  \edef\oldtheequation{\theequation}%
  \setcounter{equation}{0}%
  \def\theequation{\oldtheequation\alph{equation}}}

\def\endsubequations{\setcounter{equation}{\@savedequation}%
  \@stequation=\expandafter{\@savedtheequation}%
  \edef\theequation{\the\@stequation}\global\@ignoretrue
  \vspace*{-12pt} \\}
\def\bs{\begin{subequations}}
\def\es{\end{subequations}}
\relax
\def\thefootnote{\fnsymbol{footnote}}
\def\be{\begin{equation}}
\def\ee{\end{equation}}
\def\ba{\begin{eqnarray}}
\def\ea{\end{eqnarray}}

\def\ee{\end{equation}}
\def\bea{\begin{eqnarray}}
\def\eea{\end{eqnarray}}

\newcommand{\uarrw}[0]{\mathrel{
{\raise.5ex\vbox{\hrule width 1cm}\hskip-6pt\rightarrow}}}
\def\thebibliography#1{%
\vskip 0.5cm \centerline{\bf References}
\list{%
[\arabic{enumi}]}{\settowidth\labelwidth{[#1]}
\leftmargin\labelwidth
\advance\leftmargin\labelsep
\usecounter{enumi}}
\def\newblock{\hskip .11em plus .33em minus .07em}
\sloppy\clubpenalty4000\widowpenalty4000
\sfcode`\.=1000\relax}

\renewcommand{\theequation}{\arabic{section}.\arabic{equation}}

\renewcommand{\section}{\setcounter{equation}{0}\@startsection%
{section}{1}{0mm}{-\baselineskip}{0.5\baselineskip}%
{\normalfont\normalsize\bfseries}}

\renewcommand{\subsection}{\@startsection%
{subsection}{2}{0mm}{-\baselineskip}{0.5\baselineskip}%
{\normalfont\normalsize\slshape}}

\renewcommand{\subsubsection}{\@startsection%
{subsubsection}{2}{0mm}{-\baselineskip}{0.5\baselineskip}%
{\normalfont\normalsize\slshape}}

\begin{document}
%
%
\renewcommand{\theequation}{\arabic{section}.\arabic{equation}}
\begin{titlepage}
\begin{flushright}
hep-th/0402126 
\end{flushright}
\begin{centering}
\vspace{1.0in}
\boldmath
{ 
\bf{ \large Naturally Time Dependent Cosmological Constant} \\
\vspace{0.2cm}
\bf{ \large in} \\
\vspace{0.2cm}
\bf{ \large String Theory}   
}
\\
\unboldmath
\vspace{1.7 cm}
{\bf Andrea Gregori}$^1$ \\
\medskip
\vspace{3.2cm}
{\bf Abstract}\\
\vspace{.2in}
In the light of the proposal of Ref.~\cite{estring}, we discuss
in detail the issue of the cosmological constant,
explaining how can string theory naturally predict the value
which is experimentally observed, without low-energy supersymmetry.
\end{centering}

\vspace{4cm}

\hrule width 6.7cm
\noindent
$^1$e-mail: agregori@physik.hu-berlin.de

\end{titlepage}
\newpage
\setcounter{footnote}{0}
\renewcommand{\thefootnote}{\arabic{footnote}}

\tableofcontents

\vspace{1.5cm} 

\noindent

\section{Introduction}
\label{intro}

One of the most puzzling problems of modern  theoretical physics
is the explanation of the existence of a non-vanishing 
``vacuum energy'' in the Universe. In the recent years the non-vanishing
of this quantity  has been experimentally confirmed \cite{debern,melch}. The
cosmological constant has been measured with a certain degree of accuracy,
and it turns out to be $\; \sim 10^{-122}$ in ${\rm M}^2_{\rm P}$ units,
of the order of the squared inverse age of the 
Universe ($\sim 10^{61} \, {\rm M}^{-1}_{\rm P}$). The fact that the value of 
a physical cosmological parameter of the dimension of an energy squared, 
and therefore an inverse time squared, turns out to be precisely given by
the inverse square of the age of the Universe, induces to suspect
that it could be a manifestation of quantum mechanics at a cosmological level.
A scenario able to justify this value should then by consistency
In such a  framework, the value of the cosmological
constant should in fact correspond to the
minimal energy fluctuation ``out of the vacuum'' allowed for a quantum system
aged as our Universe, and it should find its explanation in a quantum
gravity framework. This however means that we should consider the
space-time itself as an ``entity'' extended up to the present horizon,
whose existence already by itself constitutes a quantum system, 
even before taking into account possible matter and/or radiation inside it. 
From this point of view, the empty space-time 
is not ``empty'', even when any contribution of matter and
radiation has been subtracted. This concept of vacuum
state is natural in string theory, basically a ``quantum theory
of the space-time (coordinates)''.
Indeed, string theory provides a configuration of space-time (the target
space coordinates) and, for any configuration,
possesses a vacuum state with a definite energy.
However, if, once the string configuration has been specified,
the computation of the vacuum energy is in principle unambiguous,
what seems to be not so natural is to explain in a string framework
the actual value of the cosmological constant, i.e. to produce the
``right'' string configuration. In particular, what appears very hard
is to couple a small value of this parameter with a heavy supersymmetry
breaking. In this note, we will discuss how the present experimental value
of the cosmological constant can be obtained in a natural way from string 
theory, if the latter is provided with an appropriate interpretation
of the target space coordinates. 
The point of view we are going to consider is the one of Ref~\cite{estring},
that we will now explain more in detail.
The usual way string theory is interpreted is in terms of a map
from world sheet to target space coordinates, in which the latter describe 
points in the space-time. The specific point of this map inside the space-time
has no special meaning: space-time is infinitely extended and possesses
no privileged point to be considered as the absolute ``origin''
of a physical frame better than any other one. This means that
the system possesses invariance under space and time translations.
As discussed in Ref.~\cite{estring}, one may wonder whether this is actually 
an unavoidable assumption, or whether with this we are requiring too much from
our theory. After all, what we can observe at any time
is the Universe just up to the horizon. 
Objects placed beyond the horizon can not influence our present
observations, and there is no deep ground to require, for a theory
supposed to describe the world we observe,
to know, at any given time, also about what happens outside our horizon.
At any time the amount of space we can observe is bounded by the horizon
set by the distance covered by light since the origin of the Universe,
and therefore has a radius given, in units of the speed of light,
by the age of the Universe.
For such a  theory, at any given time the space is then compact.

The hypothesis we make is that the string target space coordinates
describe this kind of ``compactification''. 
However, unlike an ordinary compactification of coordinates, e.g. onto circles,
here the compact space \emph{is}, at any time, \emph{the space}. This means
that it has not to be thought of as something embedded into something else.
What is infinite in our case is the span, the range of values,
of the string coordinates, which a priori are allowed to
take, during the history of the Universe, values in an infinitely extended
four-dimensional space. The physical space-time is however always finite,
and there is no invariance under translation of space and/or time
coordinates, because in such a space it is not immaterial the location 
of a point, with respect to the boundary of the space.
The  translation of a coordinate, $x \, \equiv \, (x_0,x_1,x_2,x_3) \,
\to x + a$, amounts to going closer or farther
from the horizon, and this  is not a symmetry.
Moreover, there is no embedding space-time in which to translate the full
system. The coordinates are therefore ``absolute''. We talk in this case of
``absolute space-time''.

\section{The cosmological constant}
\label{cosmo}

We will now discuss how, in this new interpretation, 
the present-day small value of the cosmological constant,
corresponding to the inverse of the (squared) Hubble constant,
arises as a consequence of the absence of low-energy supersymmetry.
The first consideration in order is: what do we have to intend for 
the cosmological constant? Historically introduced
as a ``constant'' term in the Einstein's action, it can be shown to
correspond to the energy of the ``vacuum''. 
There is some arbitrariness in the interpretation
of this number, in the sense that an analogous contribution
to this  effective parameter may be provided, through
the energy-momentum tensor term,  by matter and fields.
Therefore, it is possible to think that this term receives contribution  
from ``dark'' energy or matter. What we consider here is 
the pure \emph{energy of the vacuum}, namely, the energy of space-time
itself, once the contribution of any kind of matter and radiation
has been subtracted. Therefore, there is no ``microscopical'' field
description of processes contributing to this quantity: this is
a pure quantum effect. As such, the ``total'' value of the cosmological
constant comes from the string computation of the vacuum energy.
Differently from what happens in field theory,
in string theory the ``ground'' value is unambiguously determined,
once the particular ``string vacuum'' is specified.
Normally, in a situation of supersymmetry broken at the string/Planck scale,
the value produced by string theory
is too large, unless one introduces some mechanism enabling to lower
the string scale, or to make the value of the cosmological constant
independent of it, such as the existence of low-energy supersymmetry
in some sector of the theory (e.g. gravity sector) in which experiments
have been so far less sensitive.
The situation we want to consider is however the one of an intrinsically
non-perturbative ``M-theory'' vacuum, 
with supersymmetry ``hardly'' broken at the
Planck scale, as described in Ref.~\cite{estring}. 
Namely, the mass gap $\Delta m$ between particles, fields, and
their supersymmetric partners (including the mass difference between
graviton and gravitino) is of order one in Planck units, and 
the coupling of the theory in four dimensions is also of order one,
as the volume of the internal coordinates.
Naively, one would say that, unless miraculous non-perturbative
cancellations occur, these conditions lead to a cosmological
constant of order one. We will however see that in our scenario
this is not the case. 
The crucial point is how do we compare the string calculation with the 
``constant'' parameter in the Einstein's effective action.
Let's for the moment consider the case of a perturbative string vacuum: 
how to deal with the non-perturbative scenario corresponding to the actual 
situation of interest for us will be discussed later. 
Traditionally, parameters computed on the string side (as is eq.~\ref{zeta}),
are to be compared with the \underline{integrand} appearing in the effective 
action.
In other words, as it is usually defined, the string calculation 
computes ``densities'', values per unit of volume. The reason 
is that, in an infinitely extended space-time, there is a 
``gauge'' freedom, corresponding to the invariance under space-time
translations. There is therefore a redundancy in any calculation, 
related to the fact that any quantity computed at the point 
``$\vec{x}$'' is the same as at the point ``$\vec{x} + \vec{a}$''.
In order to get rid of the ``over-counting'' due to this symmetry,
one normalizes the computations by ``fixing the gauge'', i.e.
dividing by the volume of the ``orbit'' of this symmetry $\equiv$ the volume
of the space-time itself. Actually, since it is not possible to perform
computations with a strictly infinite space-time, multiplying and dividing
by infinity being a meaningless operation, the result
is normally obtained through a procedure of ``regularization'' of the
infinite. Namely, one works with a space-time of volume $V$, supposed
very big but anyway finite, and then takes the limit  $V \to \infty$.
In this kind of regularization, the volume of the space
of translations is assumed to be $V$, and it is precisely the
division by $V$ what at the end tells us that we have computed a density.
In any such computation this normalization is implicitly assumed. Notice
also that, in this approach, it is also implicitly assumed that, 
even for finite volume, T-duality of the space-time coordinates
remains broken, something necessary in order to be allowed to speak of
an effective action, for a universe of ``large'' volume.

In our case however, we do not assume the invariance under translations
to be preserved for compact space-time, and indeed it is not:
the situation we are describing is not the one of a 
``compactification'': for us space-time is ``absolute'', is extended
up to the ``horizon'', which sets a volume that, with some
simplification, can be thought to be represented
through a certain boundary value of the string target coordinates,
$\vec{X}$, identifying a ``radius'' of compactification. 
A translation of a point inside this space,
$\vec{x} \to \vec{x} + \vec{a}$ is not a symmetry,
being $\vec{X}$ fixed. On the other hand, a translation of the 
boundary value of the target space string coordinate $\vec{X}$
represents an evolution of the Universe, not a symmetry of the present-day
effective theory: there is no ``outside'' space
in which the coordinate $\vec{X}$ is embedded \footnote{In principle, 
one could think to translate all the points
of the ``segment'' $[\vec{0}, \vec{X}]$, namely, also the origin.
In this case, a translation would be a symmetry, and we would recover
the usual result. The point is that, 
in order to translate also the origin, which in our set up is the origin of 
the history of the Universe, and of ``time'' itself, we would need an
``embedding history'', a ``space-time'', and in particular
a ``time'', larger than, and containing, our one, 
existing therefore before and after it. 
But this is precisely what we have excluded in our basic hypothesis. 
For our theory, space-time \emph{is} the one we observe.}! 
In this framework, the volume of the group
of translations is not $V$: simply this symmetry does not exist at all.
There is therefore no over-counting, and what we compute is not a density,
but a global value. In other words, what in the traditional interpretation
is a density, the value of a quantity at a certain point of space-time
of the present-day Universe, in our case turns out to be a density in the 
``space'' of the whole history of the Universe, the (global) value of such 
quantity at a certain point of its history.
In the specific case of the cosmological constant, we have that the usual
dependence of the cosmological constant density on the points $\vec{x}$
of space-time gets promoted to a dependence on the
points labeling the history of the Universe, $\vec{X}$:
$\Lambda (\vec{x}) \to \Lambda (\vec{X}))$. 
In order to see this, we will have to discuss some subtleties related to the 
notion of effective action, something we will do in a moment.
Notice that, in this interpretation of string coordinates, there is
no ``good'' limit $V \to \infty$, if for ``$\infty$'' we intend the
ordinary situation in which there is invariance under translations.
This symmetry appears in fact only strictly at the limit. The
volume of translations is some kind of ``delta-function'' supported
at infinity. On the other hand, this is not a problem in our picture:
infinite space-time does not really belong to the history of the ``universe''
\footnote{Namely, of the ``observed universe'', the one that matters for
the theory.}, for which the horizon, although increasing, 
will always be finite.

\subsection{\sl On the side of the effective action}
\label{ea}

In this set up, it is not a priori obvious that one can even speak of
an effective action to compare with: in order for this to have any
meaning, T-duality along the space-time coordinates
must be broken. Only in this case it is possible
to speak of ``large'' Universe and ``expansion'', as something
distinguished from the ``contraction''. The scenario we have in mind is 
therefore the one depicted in Ref.~\cite{estring}, in which
T-duality is indeed broken, and it is possible to talk about an effective
action, in which the light modes of the theory, namely those whose mass
is below the Planck scale, move in a space-time frame 
of coordinates larger than the Planck length.
In this picture, as usual heavier modes are integrated out and their
existence manifests itself only through their contribution to the 
parameters of the effective theory. Such a description is possible
at any time of the evolution of the Universe after the breaking of T-duality
along the space-time coordinates,
a condition we will from now on implicitly assume to be realized.
At any point in the evolution, labeled by a boundary value ``$\vec{X}$''
in the string space of above, there is therefore an effective action
for the light modes:
\be
S_{\vec{X}} \, = \, 
\int_{[\vec{0},\vec{X}]} d^4 x \left( R + \Lambda + \ldots \right)
\, ,
\label{s}
\ee
where the integration region is a ``ball'' centered on the
observer and bounded by the horizon. The r.h.s. of eq.~\ref{s} has been
converted to the Einstein frame, where lengths are measured in Planck units
\footnote{\label{notefb}To the purpose of this
discussion, we can safely consider a flat metric: $\sqrt{-g} =1$.
The corrections provided by matter and the cosmological constant itself
are small in Planck units, and are therefore of second order.}.
This is what in our set up substitutes the traditional effective action.
For what matters the local physics there is no much difference, as the
contribution of the boundary can safely be neglected at our time of 
``very large'' Universe. 
When instead we want to look at ``cosmological'' phenomena, such as 
the cosmological constant, or the cosmological evolution of couplings,
masses and, in general, the ``fundamental constants'', the boundary
enters heavily in the game. Let's concentrate on the second term
of the r.h.s. of \ref{s}, namely:
\be
S_{\vec{X}}[\Lambda] \, \equiv \, \int_{[\vec{0},\vec{X}]} \Lambda \,d^4 x  
\, . 
\label{sl}
\ee  
On the side of the effective action, 
the integration is performed 
on a domain extended as the Universe up to the present-day horizon.
The effective action of  local physics is therefore 
obtained by extrapolating the local physics, the one
valid at the point where the observer is located, also to points 
located at a non-vanishing distance in space, and time, from the observer.
This is possible, because we know that the Universe
is extended (at least) up to the present-day horizon, 
and therefore we can think at this ``world'' at any time $t$ up to the present
age of the Universe, that we indicate with ${\cal T}$, 
although we cannot observe it all simultaneously\footnote{The time $t$ runs 
effectively, in Planck units, in the interval  $1 \leq t \leq {\cal T}$.}. 
On the effective theory side we have therefore
a space-density of cosmological constant $\Lambda(t)$ such that: 
\be
S_{\vec{X}}[\Lambda] \, \approx \, \int_{[\vec{0},\vec{X}]}\Lambda(t) d^4 x 
 \, .
\label{lt}
\ee
The value to be compared with the string output is therefore:
\be
\lambda(t= {\cal T}) \, \equiv \, 
\int_{[\vec{0},\vec{X}]_{t = \cal{T}}} \, \Lambda(t={\cal T}) \, d^3 x 
\; \approx \, {\cal T}^3 \Lambda ({\cal T}) \, .
\label{l3}
\ee

\subsection{\sl On the string side}
\label{ss}

Let's consider the computation of the ``vacuum energy'' in a
string vacuum of interest for our problem. Indeed, 
the physical situation we have in mind, the one described in 
Ref.~\cite{estring}, corresponds to a non-perturbative
vacuum, in which the internal coordinates are of the order of the Planck scale.
As we there observed, even when all
the (M-theory) internal coordinates, including the one playing the role
of coupling, are at the Planck scale, and therefore we are in a strong
coupling regime, it is nevertheless possible, whenever there are extended 
space coordinates, to map 
to a (dual) perturbative string vacuum in  lower dimensions. The latter 
is obtained by trading one space coordinate for
the internal coordinate playing the role of (higher dimensional) coupling
\footnote{If the (target) space of the 
theory was given in terms of a simple factorization of coordinates, 
this procedure would be as simple as rigorous.
In practice, however, we have no reason to expect such a naive situation,
and, as discussed in Ref.~\cite{estring}, there are good reasons to expect
that this is not the case. If the external space is not factorized out,
but it is in some way ``embedded'' in the whole space, that looks more
like a ``fibration'', it is not possible 
to simply trade one external coordinate for the coupling of a lower
dimensional theory: the coupling cannot be singled out. However, 
as long as we are interested just in the order of magnitude of the
cosmological constant, we do not expect the corrections due to
the embedding of the coordinates into a more complicated geometry
to be so dramatic, and the above arguments are legitimate. 
For analogous reasons, in the following we will consider the string living
in a flat background. Strictly this is not the physical case,
precisely because the non-vanishing cosmological constant
induces a curvature. However, as we already pointed out in the footnote
\ref{notefb}, such corrections to the flat background are of the order
of the cosmological constant. Therefore, they contribute only at
the second order to the determination of this quantity, and can be neglected
for the purposes of the present discussion.}. 
With good approximation the value of the vacuum energy is therefore obtained  
with a one-loop computation. Since we ``integrate over''
all space coordinates, the value obtained in this way is not the cosmological
constant of a lower dimensional theory: it is indeed the quantity
we are looking for. The reason is that the quantity we compute
is not a density but a global value, 
referred to a given point in the history of space-time, 
no matter of the role played
by the coordinates (whether coupling, or internal, or ``external'' space
coordinates) in the specific string approach we are considering. 
Thanks to this passage through a dual picture, we are therefore 
able to deal with the problem of the cosmological constant 
in a string vacuum with coupling of order one, and a (four dimensional) 
string scale of the same order as the Planck scale. 
The details of the dual realization of the actual physical configuration
are irrelevant to the purpose of the present discussion:
it is sufficient to know that \emph{all} the ground mass gaps are of order one
to conclude that the result is of order one.
In order to understand this, consider for concreteness
a ``generic'' closed string, heterotic-like realization 
of the string configuration. The results we want to extract are however to 
be thought in a string duality-invariant perspective, for which the choice 
of a specific vacuum is just a matter of convenience.
The ``cosmological term'' reads then: 
\be
S_{\vec{X}}[\Lambda] 
\, \sim \, {\cal Z} \approx \int_{\cal F} \sum_i \, (-1)^Q \, 
q^{P^{(i)}_{\rm L}} \bar{q}^{P^{(i)}_{\rm R}} \, ,
\label{zeta}
\ee  
where ${\cal Z}$ is the one-loop partition function.
As usual, $q = {\rm e}^{2 \, \pi \, {\rm i} \tau}$, ${\cal F}$
indicates the fundamental domain of integration of the torus complex
coordinate $\tau$ and $Q$ indicates the supersymmetry charge. 
The string computation tells us that, under 
the condition of mass gap between target-space ``bosons''
and ``fermions'' of order one in Planck units, $\Delta P \sim 1$,
in the \emph{light-cone gauge} the partition function is of order one:
\be
{\cal Z} \sim {\cal O}(1) \, .
\label{lambda}
\ee
In the infinite volume case, when translation and
reparametrization invariance is unbroken, the value obtained in this way
can be directly compared with the density parameter appearing in the integrand
of the effective action, eq.~\ref{sl}, to conclude that 
$\Lambda \sim {\cal O}(1)$. In our case, however, 
besides the space-time volume
factor we loose owing to the missing translational invariance, we have also
to take into account the fact that now also reparametrization is not
anymore an invariance of our theory. This means that, although this 
remains a basic invariance of the underlying string construction, 
the comparison with the effective action must take into account 
the real ``state'' of the coordinates under which a certain result has been 
obtained. The map between world-sheet and target-space coordinates is
forcedly degenerate, and the reparametrization invariance tells us that
we have the freedom to rescale coordinates in such a way to, roughly
speaking, ``identify'' two target-space coordinates with the 
world-sheet ones. String amplitudes are then obtained by integrating
over these coordinates up to their ``horizon'', in this case
corresponding to the string world-sheet size. This means that we have
shrunk a dimension-two subspace of space-time to the string proper size.
The real effective action to compare with is therefore the one in which
space-time has also been shrunk in this way. We can adjust for this
``asymmetry'' of space-time by simply switching on an inverse Jacobian
for this rescaling of coordinates on the string side, thereby multiplying
the string result by a space-time ``sub-volume'' $V_2$. 
This volume is basically obtained by squaring the Jacobian
corresponding to the boost of the time interval from the string proper length
to the age of the Universe. If $x_0$ is the world sheet variable along 
the world sheet time coordinate, running in an interval of length 
$\ell_{\rm S}$, the target space time coordinate is 
$t \sim {\cal T} \times x_0 / \ell_{\rm S}$, where 
${\cal T}$ is the age of the Universe.
The Jacobian under consideration is 
$\vert \partial t / \partial x_0 \vert$, and we obtain:
\be
\lambda(t)\vert_{(string~frame)} \, = \, 
\left\vert  {\partial t \over \partial x_0} \right\vert^2 \times {\cal Z}
~~ \sim \, {\cal T}_{\rm S}^2 \times  {\cal Z}
\, ,
\label{lsf}
\ee
where ${\cal T}_{\rm S}$ is the age of the Universe in string units.
The final step of any string computation, required in order 
to compare with the effective action, is to convert the
computation performed in the string frame into string-string duality
invariant variables, i.e. to the Einstein frame. In this frame, the proper
length is not the one of the string,$\ell_{\rm S}$, but corresponds to the 
Planck scale, $\ell_{\rm P} ~ \equiv {1 \big/ {\rm M}_{\rm P}}$ in $c=\hbar=1$
units. Expression (\ref{lsf}) is given in units of the string mass scale,
${\rm M}_{\rm S} \equiv {1 \big/ \ell_{\rm S}}$.
However, for what we said, despite the passage through a perturbative vacuum,
this expression accounts for the actual value of a four-dimensional theory
with coupling of order one, in which therefore the string scale can be 
considered equivalent to the Planck scale, $\ell_{\rm S} \equiv \ell_{\rm P}$.
The Jacobian $\left\vert  {\partial t \over \partial \ell_0} \right\vert$ 
turns out therefore to be the age of the Universe ${\cal T}$, 
as measured in Planck units: ${\cal T}_{\rm S} \approx {\cal T}$. 
As a result, we obtain\footnote{Notice that the integration
over the world-sheet coordinate $\tau \equiv x_0 + {\rm i} x_1$, 
implied by the one-loop computation of ${\cal Z}$
responsible for this result, does not correspond to an integration
over the history of the Universe: this would instead correspond
to an integration over the values of the Jacobian, 
$\vert \partial t / \partial x_0 \vert = {\cal T}$.}:
\be
\lambda(t)\vert_{string} \, \sim \, 
{1 \over {\cal T}} \times {\cal T}^2 \times
{\cal Z} \, \approx \, {1 \over {\cal T}} \times {\cal T}^2 \times
{\cal O}(1) \, .
\label{lstring}
\ee  
The factor ${\cal T}^2$ is precisely the two-volume factor $V_2$ we have just
discussed. The factor ${1 / {\cal T}}$ comes instead from the
reintroduction of a normalization of the string computation to match with
the fact that on the effective action we did not integrate over the time
coordinate, therefore we don't have to compare with an integral value
along the time.
The comparison of the string output with the effective action gives therefore:
\be
\lambda(t= {\cal T}) ~ \, \approx \, {\cal T}^3 \times \Lambda({\cal T}) 
~ \, \approx \,
{\cal T} \times  {\cal O}(1) \, ,
\label{lt3}
\ee
where the second term comes from the integral of the effective action
over the space coordinates, eq.~\ref{l3}, while the third term
is the string result. From  \ref{lt3} one obtains immediately:
\be
\Lambda ({\cal T}) \,  \sim \, {1 \over {\cal T}^2} \, \approx \, 10^{-122} \,
{\rm M}^2_{\rm P} \, . 
\ee
We stress that  it is precisely the ``built in'' failure of invariance 
under time translations, implied by our starting point of considering
space-time up to the horizon as the ``whole'' space-time of the theory,
what is the responsible for the time dependence of this parameter.
This is a key point of our set up.
It is precisely this change of perspective what allows us
to obtain time-dependent parameters without  time-dependent 
vacuum expectation values of fields.
In our framework, not only the cosmological constant naturally acquires 
a time dependence, but also curvature terms, and, as we saw in 
Ref.~\cite{estring}, masses and couplings. With the interpretation of the
string space-time compactifications as describing ``points'' in the history
of the Universe, the lack of space-time translations invariance
is then precisely equivalent to the statement of time dependence 
of ``fundamental constants''.

\section{\bf Comments}
\label{comments}

In the previous section we have discussed the comparison between 
string theory and a parameter of the effective action, 
the cosmological ``constant'', in which the later was allowed to carry
an explicit dependence on time. 
If one takes the traditional approach and considers the space-time
infinitely extended, then the cosmological evolution must be 
explicitly introduced in the effective action, and this is the only 
choice. If instead the domain of integration is supposed to be bound by the 
horizon, as in our case, it is also possible to
take a different point of view, and consider the parameters of the effective
action as constant inside the domain of integration. 
The two approaches turn out to be equivalent,
because in the second case the time dependence is anyway introduced,
through the dependence on time of the boundary, namely of the distance
of the horizon from the observer. This means that, 
at any time of the evolution of the Universe,
we will have to compare string theory with a different effective action,
differing not only in the extension of the domain of integration but
also in the value of the effective parameters. This second approach
is in some sense more ``democratic'', in what it does not 
artificially single out the evolution only along the coordinate 
we call ``time''.  

Indeed, when the horizon is finite
both approaches are equivalent, because they differ by a 
volume factor of the time, corresponding to the integration 
one has to perform, in the second case, also over this coordinate,
in order to compare with the effective action corresponding to the 
actual configuration of the Universe. This volume factor cancels
against an analogous factor present in the definition of densities.
In the first case, where the time dependence of the cosmological constant 
is explicit, in order to compare with the time-dependent parameter of the 
effective action  we don't have to integrate over the time coordinate;  
there is on the other hand also no translation invariance 
along the time coordinate, and therefore we also miss a volume factor of 
this coordinate in the normalization of the density. 
In order to see the equivalence of the two approaches, let's see
here how the comparison goes in the second case, namely when we don't
introduce an explicit time dependence in the cosmological constant.
In this case eq. \ref{lt} becomes:
\be
S_{\vec{X}}[\Lambda] \, \approx \, \int_{[\vec{0},\vec{X}]}  
\Lambda\left( {\cal T} \right) \,  d^4 x \, , 
\label{sltau}
\ee
where $d^4x = dt \, dx_1 dx_2 dx_3$. Notice that, in the integrand on the 
r.h.s., $\Lambda$ does not depend on the ``bulk'' time variable $t$ but on the 
boundary value ${\cal T}$. It is therefore a constant in the domain of 
integration. We have on the other hand a series of different effective 
actions, one for each point of the evolution of the ``boundary''coordinate 
${\cal T}$, each one with a different parameter $\Lambda$.
${\cal T}$ can therefore be considered here as a parameter labeling the 
various effective actions, one per each size of the horizon.

As opposite to the previous case,
now what we have to compare with the string result 
is the global value of this effective cosmological term, i.e. the full
expression  \ref{sltau}, which includes also the integration
over the time coordinate $t$. Instead of expression \ref{l3}, we have here:
\be
\int \lambda (t = {\cal T}) \, d t \, = \, \int_{[0,\vec{X}]} 
\Lambda ({\cal T})\, d^4 x ~ \approx ~ {\cal T}^4 \Lambda ({\cal T}) \, . 
\label{l4}
\ee 
In this case in fact the effective theory is invariant 
under translation of the ``time'' coordinate $t$ running inside the region 
of integration $[0,{\cal T}]$ (being $\Lambda$ constant, in order to well
define translations also close to the boundary, we can think to identify
the points $0$ and ${\cal T}$, and introduce periodic boundary conditions). 
The comparison with the string computation performed in the light-cone gauge
gives then: 
\be
S_{\vec{X}}[\Lambda] \, \approx \, \int_0^1 d^2 x \,
\int_0^{\cal T} d^2 x \, \Lambda({\cal T}) 
\, = \, {\cal T}^2 \times \Lambda ({\cal T}) \,  \Leftrightarrow 
\, {\cal O}(1) \, ,
\label{lt2}
\ee 
where now we obtain a factor ${\cal T}^2$ and not just ${\cal T}$.
On the right of the symbol $\Leftrightarrow$ we quote the string output.
By switching on the Jacobian ${\cal T}^2$, accounting for the transformation to
the real size of the present day coordinates, we recover the conditions 
of the output obtained from the effective action, \ref{l4}, with all the four 
space-time coordinates extended:
\be
S_{\vec{X}}[\Lambda] \, \approx \, \int_0^{\cal T} d^4 x \Lambda({\cal T}) 
\, = \, {\cal T}^4 \times \Lambda ({\cal T}) \,  \Leftrightarrow 
\, {\cal T}^2 \times {\cal O}(1) \, .
\label{lt4}
\ee 
The result is in any case the same: 
$\Lambda({\cal T}) \sim 1 / {\cal T}^2$.

\section{\bf Conclusions}
\label{conclusions}

In this work we have discussed how a change of point of view
allows to justify in a natural way the observed value of the cosmological
constant in string theory. One of the key points of our set up, proposed in 
Ref.~\cite{estring}, is that the ``effective'' space-time of the theory has, 
at any point of its history, a finite volume, the one of the region bounded by
the horizon of observation. In this framework, there is
no invariance under space-time translations: space-time is ``absolute''.
When considered as a quantum system, this ``universe'' turns out
to possess a minimal energy gap, accounted for by the so called
``cosmological constant'' (as we discussed in Ref.~\cite{estring},
masses themselves can be interpreted somehow as the effect of the ``inertia'', 
or resistance, of the system to the attempts to change its configuration 
by displacing an object from a geodesic to another one).
In absence of low-energy supersymmetry,
the value of the cosmological constant turns out to correspond to the one 
we would obtain by considering that this parameter is a measure 
of the (square of the) ``energy gap'' of the Universe within the time of 
its existence, as derived from the Heisenberg's Uncertainty Principle:
\be
\Delta E \, \Delta t \, \geq \, 1 \, ,
\label{dedt}
\ee
if we use $\Delta t = {\cal T} \, \equiv$ the age of the Universe. 
This relation has the following interpretation: since the ``time'' length
of our observable Universe is finite, the vacuum energy cannot 
vanish: it must have a gap given by (\ref{dedt}).
The cosmological constant, which has the dimension of an energy squared, 
must then correspond to the square of this quantity. In other words,
it originates as  a pure quantum effect: since it
is not protected by any symmetry, it is 
naturally generated by ``quantum corrections'', and its size
is set by the natural cut-off of the Universe.
Notice that, in the framework of our proposal,
a relation between the value of the cosmological constant and
the rate of expansion of the Universe is established at a quantum level. 
This relation is however based on the fact that both the cosmological 
``constant'' and the rate of expansion depend on the point in the history
of the Universe, i.e. on its age. There is no ``direct'' relation of the 
cosmological constant to the Hubble parameter, independently on the
age of the Universe: this comes rather as a consequence of the fact that the
cosmological constant, the rate of expansion of the Universe and
the time dependence of masses and couplings are all related
to the expansion of the horizon, the latter being a consequence
of the existence of massless modes (photon, graviton). 
What happens is that: 1) the expansion of the
horizon generates a time dependence of the cosmological constant and
masses; 2) a decrease of masses produces in turn the expansion of 
the Universe. This phenomenon is basically due to the effective 
repulsive force felt by matter in a system
in which masses decrease with time (Ref.~\cite{estring}).
As we discussed in Ref.~\cite{estring}, the non-homogeneous mass decrease,
in particular the different time-scaling of the mass of proton and  
electron, leads to a shift in the wavelength of emitted radiation 
corresponding to the one recently observed in ancient quasars 
(Refs.~\cite{dfw,wetal,metal,alpha}), an effect sometimes attributed
to the ``time-dependence of the coupling $\alpha$'' (also in  
our scenario the coupling varies with time, but in our case
its variation is not the (main) responsible for the shift in the spectra).  
On the other hand, the change with time of masses and couplings
is not in contradiction with the bounds coming from the Oklo natural
reactor (Ref.~\cite{oklo}) or from the nucleosynthesis. 
Both in fact are bounds on cross sections, whose variation depends
not just on a single scale but on the overall effect of all the mass and 
coupling variations. Only under the hypothesis of keeping fixed
all scales except from the coupling allows to convert these constraints 
into bounds on the variation of this coupling. This is not our case.

Giving up with the idea of including infinity within the ``regular''
configurations of the system, namely, considering the history of 
space-time volumes as an open set, implies that there is no non-trivial 
limit of parameters of the theory, 
as the volume $V$ of the Universe ``inside the horizon'' 
expands toward infinity. As $V \to \infty$, the parameters
of the theory, i.e. couplings, masses, and the cosmological constant,
tend either to zero or to infinity (this second is the case of the strong 
coupling and masses of string states above the Planck scale).
A non-trivial value for these parameters is instead obtained 
in the traditional approach, in which infinity is included in the
configurations of the Universe. This introduces a divergence in the (string)
theory, whose regularization leads to finite non-trivial limits for these
parameters: for instance, for the vacuum energy.

As opposite to the commom point of view, here we don't consider a space-time 
of finite volume $V$ as a sort of ``regularization of infinity'', artificially 
introduced in order to solve computational problems, while  having however 
in mind the infinite volume limit as the true, physical situation.
In our approach, a space-time of finite volume is the ``real'', 
effective physical situation at any finite time.  
This  perspective allows us to address many physical questions in a completely 
different way. As discussed in Ref.~\cite{estring}, a consequence of 
the above  mentioned ``trivialization'' of infinity is that now 
the ``geometry'' of space-time turns out to evolve toward a ``flat'' limit.    
This is in some sense the ``classical'' limit of the system, an asymptotic
configuration never realized in practice.
The scenario is therefore quite different from the usual one, in which 
cosmology is based on  a model of the Universe considered as a 
perfect fluid. Here this description cannot be applied, because the
``truncation'' of the theory at the horizon prevents from introducing
a smooth metric for a space-time extending also beyond it. 
The kinematic equations 
for the expansion of the Universe cannot be transferred in a simple way to
the present scenario, because in our case the expansion is not adiabatic.
Indeed, since the ``boundary'' of the effective universe
expands at the speed of light, its energy
is not conserved: if we start at a certain point with a certain
matter content, the energy corresponding to this ``small universe''
will be conserved under expansion, because by definition the horizon 
itself expands together with radiation (there is no radiation 
flowing out of the horizon). 
However, new matter will appear inside the horizon, and energy 
will not be conserved for this system. This effect is not balanced by 
a decrease of vacuum energy: the vacuum energy density scales in fact
like the square root of the cosmological constant, therefore
as the inverse of the age of the Universe, $\rho(E)_{\rm vac} 
\sim {1 \over {\cal T}}$, 
and the total vacuum energy within the horizon increases therefore 
as the square of the age of the Universe, $E_{\rm vac} \sim {\cal T}^2$. 
The system is therefore somehow holographic, at least for what matters
the properties of the vacuum. In this set up, this is on the other hand
not the case for the matter content and its energy. 
However, we have seen in Ref.~\cite{estring}
that matter does not feel an ``ordinary'' geometry of space-time; there is 
therefore nothing to worry about the fact that the matter energy density 
does not fit with a simple geometric volume/boundary relation.

In the interpretation proposed in Ref.~\cite{estring}, the quantization
of the string coordinates, among which is our space-time, 
turns out to be something quite
different from the ordinary \emph{local} quantization of fields and particles.
We have two ``levels'' of quantization: the local
one, concerning just the neighbor of the observer, for which it is
a good approximation to consider space-time as infinitely extended. This
leads to the ordinary quantization and an effective action
describing a ``time-independent'' physics. We have then   
the level of the quantization of space-time itself, 
leading to the cosmological evolution of the Universe. 
As opposite to the traditional approach, what we propose is  
that string theory does not provide us with an ``improved'' 
description of the ``local'' physics.
It gives something else: the evolution of the ``bubble'' inside which
we can define what we call ``space-time'', namely the frame in which
everything we observe takes place. 
In this approach, certain parameters of the ``local'' effective theory,
like masses and the cosmological constant, have to be seen as
effective terms, whose origin is 
explained by switching to the ``cosmological'' description, the 
``global'' description provided by string theory interpreted as we said.
These parameters are related to global facts, like the size of the
horizon, and therefore depend on the point $\vec{X}$ in the history
of the Universe. 

\vspace{1.5cm}
\centerline{\bf Acknowledgments}

\noindent
I thank G. Veneziano for useful observations.

\vspace{1.5cm}


\providecommand{\href}[2]{#2}\begingroup\raggedright\endgroup

\end{document}